\documentclass[a4paper,11pt]{article}

\usepackage{jheppub}
\usepackage[T1]{fontenc}
\usepackage{amssymb}
\usepackage{bm}
\usepackage{mathtools}
\usepackage{booktabs}
\usepackage{tabularx}
\usepackage{array}
\usepackage{graphicx}
\usepackage{subcaption}
\usepackage{float}
\usepackage[section]{placeins}
\graphicspath{{figures/}}
\hypersetup{hypertexnames=false}

\usepackage{orcidlink}

\newcommand{\iu}{\mathrm{i}}
\newcommand{\dd}{\mathrm{d}}
\newcommand{\wep}{w_{\mathrm{EP}}}
\newcommand{\gep}{g_{\mathrm{EP}}}
\newcommand{\nuep}{\nu_{\mathrm{EP}}}

\newcommand{\co}{\mathrm{co}}
\newcommand{\ctr}{\mathrm{ctr}}
\newcolumntype{Y}{>{\raggedright\arraybackslash}X}
\newcommand{\btzpanel}[3]{%
 \begin{subfigure}[t]{#1}
  \centering
  \includegraphics[width=\linewidth]{#2}
  \caption{#3}
 \end{subfigure}%
}

\title{\boldmath Exceptional points and near-extremal boundary response in rotating BTZ black holes}

\author[a]{Sheng Long,}
\author[b]{Yu Shi,}
\author[c]{Zhongwu Xia,}
\author[c]{Huajie Gong,}
\author[a, d,*]{Zhoujian Cao,}
\author[c,*]{Qin Tan,}
\author[c, e, *]{and Qiyuan Pan}

\affiliation[a]{School of Fundamental Physics and Mathematical Sciences, Hangzhou Institute for Advanced Study, University of Chinese Academy of Sciences,\\
No. 1 Xiangshan Branch, Hangzhou 310024, People's Republic of China}
\affiliation[b]{Department of Physics, Huaihua University, Huaihua, Hunan 418008, China}
\affiliation[c]{Hunan Normal University,\\
36 Lushan Road, Changsha, Hunan 410081, People's Republic of China}
\affiliation[d]{Department of Astronomy, Beijing Normal University, Beijing 100875, China}
\affiliation[e]{Center for Gravitation and Cosmology, College of Physical Science and Technology,\\
Yangzhou University, Yangzhou 225009, People's Republic of China}
\affiliation[*]{Corresponding authors.}

\emailAdd{zjcao@bnu.edu.cn}
\emailAdd{tanqin@hunnu.edu.cn}
\emailAdd{panqiyuan@hunnu.edu.cn}

\abstract{
We investigate whether the coalescence of long-lived quasinormal modes
enhances the boundary response of rotating BTZ black holes. Using exact
scalar solutions with mixed Robin boundary conditions and the retarded
boundary correlator, we compare frequency-domain resonances and
time-domain signals at fixed source and operator normalizations.
An exceptional point (EP) can be crossed by varying the spin alone in
a suitably selected fixed theory. Along a distinct near-extremal path
that tracks the co-rotating EP surface by tuning the scalar mass and
boundary coupling, a taller, narrower resonance coexists with a weaker
but longer-lasting double-pole impulse response. The frequency-domain
enhancement persists away from the EP and is therefore not unique to
mode coalescence. Thus, the EP determines the double-pole structure
and Jordan relaxation, but does not by itself determine the response
strength, which also depends on the physical path and excitation
protocol.
}

\keywords{Black Holes, Quasinormal Modes, Exceptional Points, AdS/CFT Correspondence, Pseudospectra}

\begin{document}
\compress
\maketitle
\raggedbottom

\section{Introduction}
\label{sec:introduction}

A perturbed black hole exhibits damped oscillations whose characteristic
frequencies and decay times are encoded in its quasinormal modes (QNMs).
These resonances are selected by absorption at the horizon together with an
outer boundary condition, and provide a way to study how the black hole
responds to disturbances
\cite{Kokkotas:1999bd,Berti:2009kk,Konoplya:2011qq}.
Their complex frequencies describe both the oscillation and the loss of the
perturbation, connecting the spectral problem to the physical process of
black-hole relaxation.

Black-hole quasinormal modes encode the characteristic oscillation and damping
timescales of the ringdown, and their frequencies can therefore be used to
infer remnant properties and test the black-hole spectrum
\cite{Dreyer2004,BertiCardosoWill2006,Giesler2019,Isi2019}. Beyond characterizing black holes within general relativity, QNM spectra also provide a natural probe of modified gravity, since changes in the background geometry or perturbation dynamics can shift both the mode frequencies and damping rates \cite{Sekhmani:2025zen,Deng:2025uvp,Liu:2023uft, Jia:2024sdk, Gong:2026yjg}.
Their observable imprint, however, also depends on mode excitation, source
coupling, and the corresponding pole residues, rather than on the damping
time alone \cite{BertiCardoso2006Excitation}.

The spectrum, however, does not by itself determine the strength of the
response.  A mode may decay slowly without being strongly excited by a given
disturbance.  Its contribution also depends on the excitation amplitude and
on the source that drives it.  In asymptotically anti-de Sitter (AdS)
spacetimes, this distinction can be expressed directly through the thermal
retarded correlator: QNM frequencies determine its pole positions, whereas
the residues determine how those poles enter the response
\cite{Horowitz:1999jd,Kovtun:2005ev,Son:2002sd}.
A narrow frequency-domain resonance and a large time-domain signal are
therefore different physical properties.

Studies of QNM excitation factors and time-domain reconstruction show explicitly
that the contribution of a mode depends not only on its complex frequency but
also on its residue and its overlap with the perturbing source
\cite{Leaver:1986gd,Sun:1990pi, Andersson:1995zk, Zhang:2013ksa, London:2014cma, Chung:2023zdq}.

The distinction becomes especially important when two modes merge.
At a second-order exceptional point (EP), the frequencies and eigenfunctions
coalesce, leaving only one independent eigenfunction.  The resulting
relaxation is no longer a sum of two independent damped oscillations:
the exponential acquires a factor linear in time.  The same defect produces
a local square-root splitting under parameter variation
\cite{Kato:1995,Heiss:2012qla,Ashida:2020dkc}.
In black-hole spectroscopy, EPs have been connected to avoided crossings and
resonant excitation
\cite{Motohashi:2024fwt,Yang:2025dbn,Oshita:2025ibu,PanossoMacedo:2025xnf}.
Massive scalar perturbations of Kerr exhibit mode exchange around a
real-parameter degeneracy \cite{Cavalcante:2024kerr}, while deformed
black-hole potentials admit exceptional lines and characteristic
pseudospectral scaling \cite{Cao:2025exceptionalline}.
The Nariai problem further illustrates that the visibility of a nearly
double-pole contribution depends on its excitation amplitude and on the
other modes present \cite{Nakamoto:2026nariai}.
Mode coalescence thus changes the structure of relaxation, but its observable
consequences still require a response calculation.

Rotation introduces a further question: what happens when mode coalescence
is accompanied by an increasingly long relaxation time?
Rotating BTZ provides an analytically accessible setting in which to address
this question.  Its scalar equation is hypergeometric, and its co- and
counter-rotating QNM sequences involve two distinct chiral thermal scales
\cite{Banados:1992gq,Cardoso:2001hn,Birmingham:2001pj}.
Only the co-rotating scale vanishes at extremality.
The spectrum and the boundary response can therefore be followed together
as this scale tends to zero.

Near-extremal rotating black holes can support long-lived or zero-damped QNMs, whose damping rates vanish toward extremality, although their observable excitation still depends on the corresponding residues and source coupling
\cite{Detweiler:1980gk,Yang:2012pj,Yang:2013uba,BertiCardoso2006Excitation,Gralla:2016sxp}.

The boundary condition supplies an additional physical control.
In the Breitenlohner--Freedman (BF) mass window, both scalar falloffs are
normalizable \cite{Breitenlohner:1982bm,Klebanov:1999tb}, allowing mixed
Robin boundary conditions
\cite{Ishibashi:2004wx,Dappiaggi:2017pbe}.
Once the source and renormalization prescription are fixed, these conditions
encode a double-trace deformation of the boundary theory
\cite{Witten:2001ua,Hartman:2006dy}.
We specify the boundary data using the Fefferman--Graham (FG) asymptotic
coefficients, with fixed source and operator normalizations
\cite{Skenderis:2002wp,Skenderis:2008dg}.
This makes it possible to distinguish changing the state of a fixed theory
from changing the theory itself, a distinction that matters when comparing
response amplitudes.

The spectral starting point is the nonrotating Robin-BTZ construction of
Wang and Yang \cite{Wang:2026robin}, which identifies the EP locus,
pole-pairing rearrangement, and associated Jordan double pole.
Zhou developed a general hypergeometric quantization and residue framework,
including a double-pole criterion and a rotating-BTZ Dirichlet benchmark
\cite{Zhou:2026doublepole}.
Dias, Sola Gil, and Santos studied rotating BTZ with double-trace boundary
conditions and explicitly distinguished fixed-theory state variations from
changes of the boundary coupling \cite{Dias:2025doubletrace}; their spectra
include a nonrotating mode merger and instability onsets.
Robin-induced spectral holonomy is also known in planar
Schwarzschild--AdS$_4$ \cite{Kinoshita:2023robin}.
These results establish the framework for our question: which features of
the rotating critical response arise from the EP, and which arise from the
thermal scale and the physical path used to approach it?

Here we address this question using the exact scalar spectrum and boundary
correlator of rotating BTZ black holes with Robin boundary conditions.
We compare frequency-domain resonances and time-domain responses along
EP and non-EP paths toward extremality, distinguishing changes of state
within a fixed boundary theory from changes of the theory itself.
Applying the same finite-duration source further allows us to compare
these paths under a common excitation.
A central distinction emerges: a taller, narrower frequency-domain
resonance can coexist with a weaker, longer-lasting double-pole impulse
response.
This provides a concrete setting in which to distinguish the effect of
mode coalescence from that of a diverging relaxation time.

Section~\ref{sec:exact} presents the exact characteristic and strict EP
criterion, and section~\ref{sec:phase} studies the chiral critical surfaces
and fixed-theory crossing.
Section~\ref{sec:holography} develops the boundary response, including the
finite-duration drive, while section~\ref{sec:pseudospectrum} presents the
regulated spectral diagnostic.
Derivations, mode-family and topology checks, and operator details are
given in the Supplemental Material.

\section{Exact Robin spectrum of rotating BTZ}
\label{sec:exact}

\subsection{Boundary data and the exact characteristic}

The rotating BTZ geometry is \cite{Banados:1992gq,Carlip:1995qv}
\begin{equation}
	\dd s^2=-N^2\dd t^2+N^{-2}\dd r^2
	+r^2(\dd\varphi+N^\varphi\dd t)^2,
	\quad
	N^2=\frac{(r^2-r_+^2)(r^2-r_-^2)}{L^2r^2},
	\quad
	N^\varphi=-\frac{r_+r_-}{Lr^2}.
	\label{eq:btz_metric}
\end{equation}
$L$ is the AdS radius and $r_+>r_-\geq0$ are the outer and inner horizon
radii.  The angular velocity and Hawking temperature are
$\Omega_H=r_-/(Lr_+)$ and
$T_H=(r_+^2-r_-^2)/(2\pi L^2r_+)$.  Absorption depends on the
corotating frequency $\omega-m\Omega_H$.

For $(\Box-\mu^2)\Phi=0$ with
$\Phi=e^{-\iu\omega t+\iu m\varphi}R(r)$, define
\begin{equation}
	a=\frac{r_-}{r_+},\qquad q=\frac{mL}{r_+},\qquad
	w=\frac{\omega L^2}{r_+},\qquad
	\mu^2L^2=\nu^2-1,
	\quad 0<\nu<1.
	\label{eq:dimensionless_controls}
\end{equation}
We take $q\geq0$ and $0\leq a<1$; the chiral labels below refer to this
Fourier and azimuthal convention.  In the open BF window, both scalar falloffs
are normalizable, so a real mixed boundary condition is admissible
\cite{Breitenlohner:1982bm,Klebanov:1999tb,Ishibashi:2004wx}.
Rotation supplies the chiral scales
$T_{\co}=(r_+-r_-)/(2\pi L^2)$ and
$T_{\ctr}=(r_++r_-)/(2\pi L^2)$; only $T_{\co}$ vanishes at extremality.
They are two factors of one radial solution, not independent Hawking
temperatures.  Indeed
$T_H=2T_{\co}T_{\ctr}/(T_{\co}+T_{\ctr})$.

With
\begin{equation}
	z=\frac{r^2-r_+^2}{r^2-r_-^2},\qquad
	h_\pm=\frac{1\pm\nu}{2},\qquad
	x_{\co}=\frac{w-q}{2(1-a)},\qquad
	x_{\ctr}=\frac{w+q}{2(1+a)},
	\label{eq:chiral_variables}
\end{equation}
the solution ingoing at $z=0$ is, up to normalization,
\begin{equation}
	R_{\rm in}=z^{-\iu(w-aq)/[2(1-a^2)]}(1-z)^{h_-}
	{}_2F_1(\mathsf a,\mathsf b;\mathsf c;z),
	\label{eq:ingoing_solution}
\end{equation}
where
$\mathsf a=h_- -\iu x_{\ctr}$,
$\mathsf b=h_- -\iu x_{\co}$, and
$\mathsf c=1-\iu(w-aq)/(1-a^2)$.
The horizon factor depends on $w-aq$, while the hypergeometric parameters keep
the co- and counter-rotating scales separate.  This distinction controls the
chiral damping hierarchy studied in section~\ref{sec:phase}.
At the AdS boundary it has the form
$R_{\rm in}\sim C_-(1-z)^{h_-}+C_+(1-z)^{h_+}$, with
\begin{align}
	C_-&=\Gamma(\mathsf c)\Gamma(\nu)A(w),
	& A(w)&=\frac{1}{
	\Gamma(h_+-\iu x_{\co})\Gamma(h_+-\iu x_{\ctr})},
	\label{eq:A_coefficient}\\
	C_+&=\Gamma(\mathsf c)\Gamma(-\nu)B(w),
	& B(w)&=\frac{1}{
	\Gamma(h_--\iu x_{\co})\Gamma(h_--\iu x_{\ctr})}.
	\label{eq:B_coefficient}
\end{align}
The future-horizon condition has fixed the relative boundary amplitudes; the
connection derivation \cite{NIST:DLMF} and normalization factors are given in the Supplemental
Material.
Because reciprocal Gamma functions are entire, the characteristic below is
entire in $w$.  Keeping $A$ and $B$ separately avoids artificial poles from
dividing one boundary coefficient by the other and permits the two endpoint
spectra and finite-Robin roots to be followed in one analytic equation.

On the real Robin branch between Dirichlet and Neumann,
$\cos\xi\,C_- -\sin\xi\,C_+=0$ with $0<\xi<\pi/2$.  Removing common
factors gives the entire characteristic
\begin{equation}
	\mathcal H(w;g,\nu,a,q)=B(w)+gA(w)=0,
	\qquad
	g=-\frac{\Gamma(\nu)}{\Gamma(-\nu)}\cot\xi>0.
	\label{eq:master_characteristic}
\end{equation}
Its $g=0$ and $g\to\infty$ limits reproduce the Neumann and Dirichlet towers,
\begin{equation}
	w_{n,s}^{(\mp)}=\sigma_s q
	-\iu(1-\sigma_s a)(1\mp\nu+2n),
	\qquad
	(\sigma_{\co},\sigma_{\ctr})=(+1,-1),
	\quad n=0,1,\ldots .
	\label{eq:alternate_endpoints}
\end{equation}
The upper and lower signs correspond respectively to the Neumann and
Dirichlet damping offsets.  At finite $g$, both chiral factors enter one Robin
condition and cannot be quantized independently.  The endpoint towers remain
useful as continuation anchors, while the EPs studied below occur at finite
positive $g$.

To identify the physical mixed-boundary coupling, we rewrite the asymptotic scalar field in the standard Fefferman--Graham (FG) normalization \cite{Skenderis:2002wp, deHaro:2000vlm, Klebanov:1999tb},
\begin{equation}
	R=\alpha_{\rm FG}r^{-(1-\nu)}
	+\beta_{\rm FG}r^{-(1+\nu)}+\cdots,
	\qquad
	\beta_{\rm FG}=\varkappa_{\rm FG}\alpha_{\rm FG}.
	\label{eq:FG_expansion}
\end{equation}
Since
$\alpha_{\rm FG}=C_-[r_+^2(1-a^2)]^{h_-}$ and
$\beta_{\rm FG}=C_+[r_+^2(1-a^2)]^{h_+}$,
\begin{equation}
	\varkappa_{\rm FG}
	=-[r_+^2(1-a^2)]^\nu
	\frac{\Gamma(-\nu)}{\Gamma(\nu)}g,
	\qquad
	\widehat\varkappa_{\mu_R}=\mu_R^{2\nu}\varkappa_{\rm FG}.
	\label{eq:g_to_FG_coupling}
\end{equation}
Holding $g$ while changing $a$ therefore does not hold the FG coupling fixed.
We use $g$ only as a Gamma-normalized Robin coordinate and state the FG
prescription whenever amplitudes or fixed theories are compared.  In the
response section the source is
$J_\varkappa=\beta_{\rm FG}-\varkappa_{\rm FG}\alpha_{\rm FG}$; fixing that
source and an operator normalization is also necessary before comparing pole
residues across states.
This distinction is central to both paths used later: the spin crossing keeps
$\varkappa_{\rm FG}$ fixed by varying $g(a)$, whereas the near-extremal
critical path tunes $g$, $\nu$, and hence the physical FG coupling.

\subsection{Strict exceptional-point criterion}

Equal frequencies alone do not establish an EP.  For the analytic radial
problem $\mathcal P(w;\boldsymbol\lambda)$ with ingoing and Robin conditions,
a second-order QNM EP requires
\begin{equation}
	\mathcal H=\partial_w\mathcal H=0,
	\qquad
	\partial_w^2\mathcal H\neq0,
	\qquad
	\partial_g\mathcal H=A\neq0,
	\qquad
	\dim\ker\mathcal P(\wep;\boldsymbol\lambda_{\rm EP})=1.
	\label{eq:EP_conditions}
\end{equation}
The derivative conditions are evaluated at
$(\wep;\gep,\nuep,a,q)$.  The second derivative excludes a higher-order
merger, and $A(\wep)\neq0$ ensures that changing $g$ unfolds the root.  Away
from resonant horizon indices, singular connection prefactors, and coincident
endpoint zeros, the horizon problem supplies a nontrivial analytic ingoing
solution unique up to normalization.  Applying the Robin functional changes
$\mathcal H$ only by a nonvanishing analytic factor.  A strict double zero
therefore has algebraic multiplicity two and a one-dimensional eigenspace:
the roots share one radial mode rather than form an ordinary crossing.
The associated function obeys
\begin{equation}
	\mathcal P(\wep)\psi_0=0,
	\qquad
	\mathcal P(\wep)\psi_1=-\partial_w\mathcal P(\wep)\psi_0,
	\label{eq:EP_Jordan_chain}
\end{equation}
which is the length-two Jordan chain underlying the time dependence in
section~\ref{sec:holography}.

At fixed $(\nu,a,q)$, a small Robin detuning unfolds the root as
\begin{equation}
	w_\pm=\wep\pm
	\left[-\frac{2A(\wep)}{\partial_w^2\mathcal H(\wep)}\delta g\right]^{1/2}
	+O(\delta g).
	\label{eq:Puiseux_local}
\end{equation}
The square root gives the local sheet exchange, while the one-dimensional
ingoing space supplies the geometric defect; neither test substitutes for the
other.  The associated function is a generalized direction, not a second
eigenmode.  Numerical residuals, representative benchmarks, endpoint checks,
and profile continuation are reported in the Supplemental Material.

\section{Chiral critical surfaces and a fixed-theory crossing}
\label{sec:phase}

\subsection{Chiral thermal scales}
\label{subsec:phase_diagram}

\begin{figure}[t]
	\centering
	\btzpanel{0.48\textwidth}{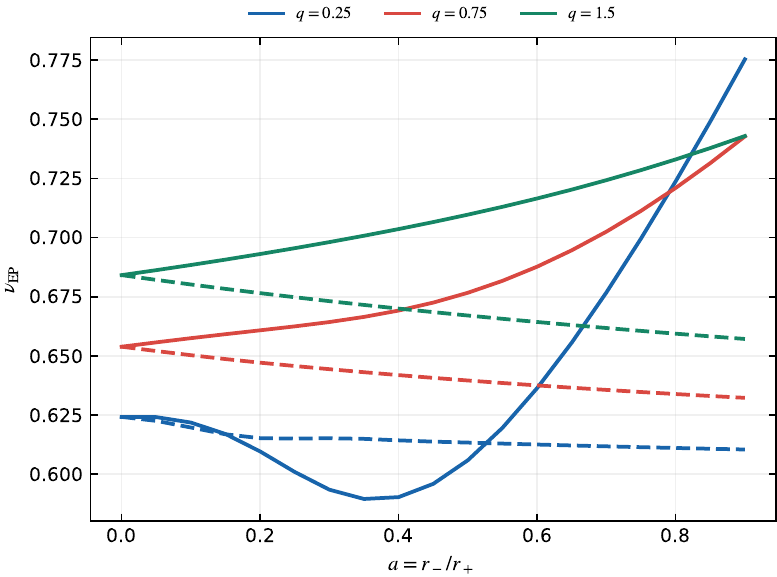}{Critical BF exponent $\nu_{\rm EP}$.}\hfill
	\btzpanel{0.48\textwidth}{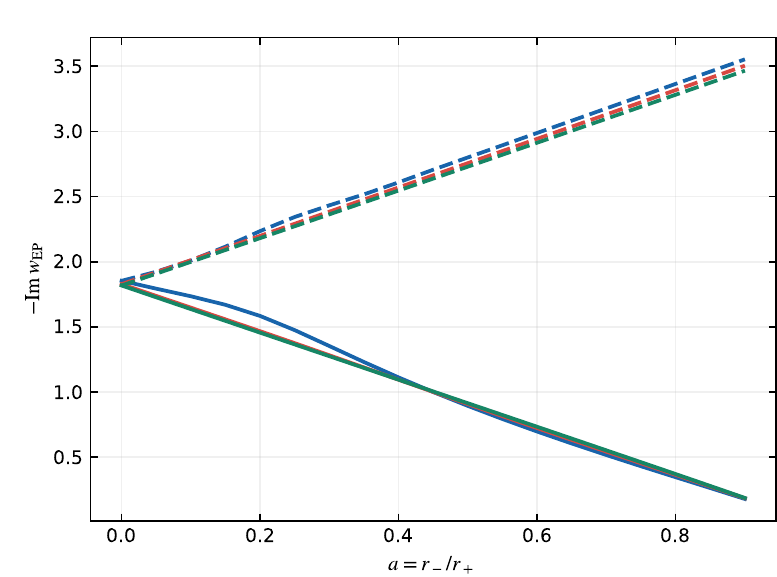}{EP damping rate $-\operatorname{Im}w_{\rm EP}$.}
	\par\medskip
	\btzpanel{0.48\textwidth}{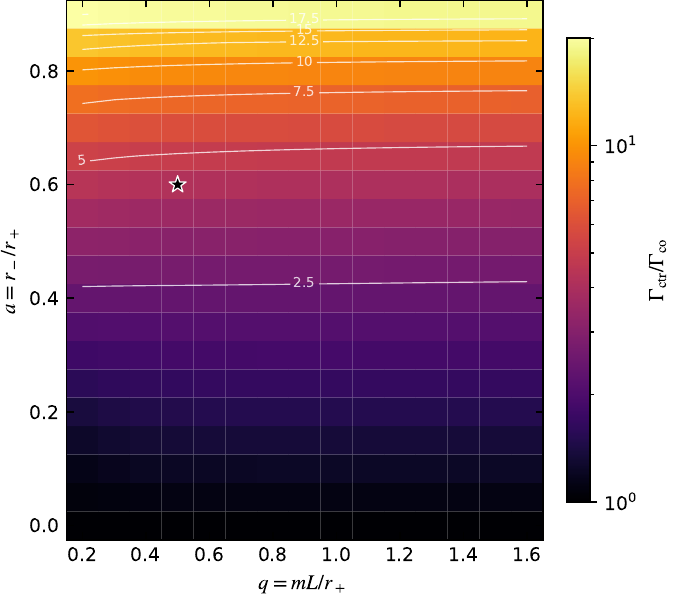}{Raw chiral damping contrast.}\hfill
	\btzpanel{0.48\textwidth}{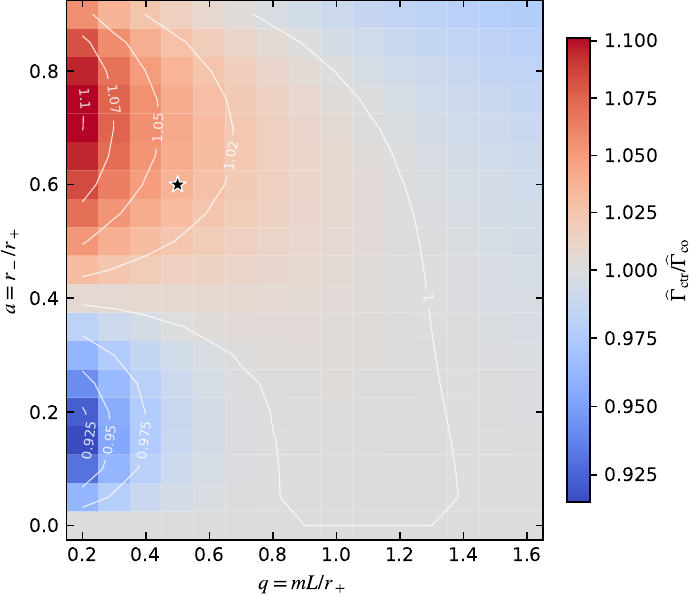}{Residual contrast after thermal scaling.}
	\caption{Least-damped chiral EP surfaces from the exact double-root
	equations.  In the upper row, solid and dashed curves denote the co- and
	counter-rotating sectors and color labels $q$.  The lower row covers
	$0.2\leq q\leq1.6$; the star marks $(a,q)=(0.6,0.5)$ and white curves are
	iso-value contours.}
	\label{fig:phase_diagram}
\end{figure}

Continuing the exact double roots over $0\leq a\leq0.9$ separates the
least-damped co- and counter-rotating critical surfaces as soon as $a\neq0$.
Figure~\ref{fig:phase_diagram} shows their critical mass data and damping.  The
two surfaces generally have different $(g_{\rm EP},\nu_{\rm EP})$, so this is
a comparison across critical theories rather than between two modes of one
fixed theory.  All displayed roots remain damped on the $g>0$ branch; other
Robin branches may admit superradiant instabilities
\cite{Dappiaggi:2017pbe}.
The curves use $q=0.25,0.5,\ldots,1.5$; the contrast maps extend the check to
15 momenta over $0.2\leq q\leq1.6$ and 19 spins.  Each cell is a separately
refined double root.  Complete control maps and their normalized residuals are
given in the Supplemental Material.

The dominant damping contrast follows the two chiral thermal scales.  Defining
\begin{equation}
	\Gamma_s=-\operatorname{Im}w_s,
	\qquad
	\widehat\Gamma_{\co}=\frac{\Gamma_{\co}}{1-a},
	\qquad
	\widehat\Gamma_{\ctr}=\frac{\Gamma_{\ctr}}{1+a},
	\label{eq:thermal_damping_normalization}
\end{equation}
gives the identity
\begin{equation}
	\frac{\Gamma_{\ctr}}{\Gamma_{\co}}
	=\frac{1+a}{1-a}
	\frac{\widehat\Gamma_{\ctr}}{\widehat\Gamma_{\co}}.
	\label{eq:thermal_damping_factorization}
\end{equation}
Across the $15\times19$ $(q,a)$ grid, the normalized ratio remains in
$0.9150$--$1.1013$, whereas the raw ratio reaches $20.0674$.  Rotation thus
sets the leading chiral hierarchy, but the residual on both sides of unity
shows that the exact critical surfaces do not collapse under a temperature
rescaling.  Their nonmonotonic mass dependence reflects the simultaneous
presence of both chiral Gamma-function arguments in
eqs.~\eqref{eq:A_coefficient}--\eqref{eq:B_coefficient}.  Complete control
maps and higher-family data are retained in the Supplemental Material.
The ratio of critical $g$ values can differ much more strongly, but
eq.~\eqref{eq:g_to_FG_coupling} makes that comparison coordinate and
scheme dependent.  It is not a ratio of fixed-theory double-trace couplings.

\subsection{A critically selected fixed-theory crossing}
\label{subsec:topology}

The same local defect can be crossed by varying the state alone if the fixed
theory is selected to pass through it.  We set
$m=L=\mu_R=1$, $r_+=2$, $q=0.5$,
$\nu=0.667339$, and $\varkappa_{\rm FG}=3.18869$, then vary $a$.
Equation~\eqref{eq:g_to_FG_coupling} determines the thermal coordinate $g(a)$
required to keep the FG coupling fixed.  The resulting path crosses the
co-rotating EP at $a_{\rm EP}=0.6$.
Here $g(a)$ changes because the conversion between the thermal basis and FG
data contains $1-a^2$.  Treating $g$ as the physical coupling would therefore
turn this fixed-theory path into a different comparison.

\begin{figure}[t]
	\centering
	\btzpanel{0.48\textwidth}{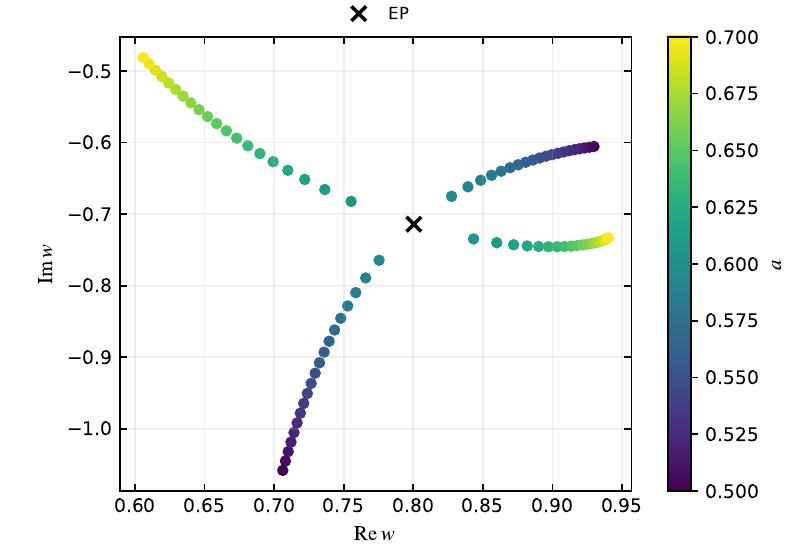}{QNM trajectories at fixed FG coupling.}\hfill
	\btzpanel{0.48\textwidth}{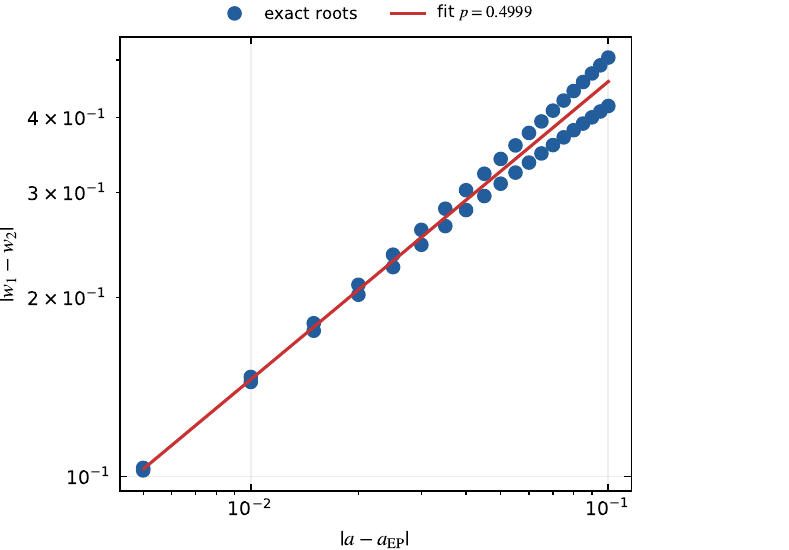}{Transverse square-root splitting.}
	\caption{Spin crossing at fixed scalar data and fixed FG coupling.  Color
	denotes $a$, and the cross marks the EP.  The one-dimensional path crosses
	the branch point but does not define a closed-loop monodromy.}
	\label{fig:fixed_theory_slice}
\end{figure}

The frequency separation follows
$|w_+-w_-|\propto|a-a_{\rm EP}|^{0.49986}$ over
$|a-0.6|\leq0.04$, verifying a transverse square-root crossing.  This path is
an existence example, not a generic consequence of changing spin: its scalar
mass and coupling were chosen from the critical point, and a generic
fixed-theory trajectory need not meet the codimension-two EP surface.
It also carries no winding number because it is a one-dimensional crossing.

Real $(g,\nu)$ loops give the complementary topology check.  One circuit
exchanges the frequency sheets and their interior-normalized radial profiles;
two circuits restore them, with signed vorticity $+1/2$ for the chosen
orientation.  The loop geometry, profile-distance definition, complex-$g$
cross-check, and numerical residuals are given in the Supplemental Material,
where they support the strict analytic criterion rather than replace it.
The interior profile normalization removes an arbitrary amplitude; the return
after two loops is not a Berry-phase measurement or a driven-state protocol.

\section{Boundary correlator and near-extremal response}
\label{sec:holography}

\subsection{Boundary normalization and the double pole}
\label{subsec:jordan_response}

We use the mixed FG source
$J_\varkappa$ and the
alternate-quantization response
$\langle O_-\rangle=-2\nu\alpha_{\rm FG}$.  The ingoing prescription and
double-trace resummation give
\cite{Son:2002sd,Skenderis:2002wp,Skenderis:2008dg,Witten:2001ua,Hartman:2006dy}
\begin{align}
	\widetilde G^{\mathrm R}_g(w,q)
	&=-2\nu\frac{A(w)}{B(w)+gA(w)},
	\nonumber\\
	G^{\mathrm R}_{\varkappa,{\rm FG}}
	&=Z_{\rm FG}\widetilde G^{\mathrm R}_g,
	\qquad
	Z_{\rm FG}=[r_+^2(1-a^2)]^{-\nu}
	\frac{\Gamma(\nu)}{\Gamma(-\nu)} .
	\label{eq:retarded_response}
\end{align}
Here and below the contact polynomial is set to zero, with no spin-dependent
finite rescaling of the source or operator.  All response figures use
$m=L=\mu_R=1$ and $r_+=2$.  These choices fix the amplitudes compared along
the paths below; they do not affect pole positions or orders.
The exact functions $A$ and $B$ in eqs.~\eqref{eq:A_coefficient} and
\eqref{eq:B_coefficient} are used throughout.  Consequently the real-axis
calculation contains the full meromorphic correlator in this channel, rather
than only the local data of one pole.  A finite operator rescaling would change
the displayed amplitudes, which is why the prescription is held fixed as the
spin and critical parameters vary.

At the EP, the double zero of the characteristic function
$\mathcal H$, together with
$A(\wep)\neq0$ in eq.~\eqref{eq:EP_conditions}, gives
\cite{Zhou:2026doublepole,Wang:2026robin}
\begin{align}
	\widetilde G^{\mathrm R}_g(w)
	&=\frac{\widetilde R_2}{(w-\wep)^2}
	+\frac{\widetilde R_1}{w-\wep}+O(1),
	\nonumber\\
	\widetilde R_2&=-4 \nu \frac{ A}{\mathcal H''},
	\qquad
	\widetilde R_1= -4 \nu \left( \frac{A'}{\mathcal H''}
	-\frac{A \mathcal H'''}{3(\mathcal H'')^2} \right),
	\label{eq:response_laurent}
\end{align}
where the derivatives are evaluated at $\wep$ and
$R_j^{\rm FG}=Z_{\rm FG}\widetilde R_j$.  Thus the geometric defect found
in section~\ref{sec:exact} is visible in this boundary channel.
The denominator is the source-free boundary condition, while the numerator
sets the coupling of that mode to the chosen operator.  Its nonzero value at
the EP rules out cancellation of the second-order pole in this channel.  The
pole order is fixed by the defect, but the coefficients in
eq.~\eqref{eq:response_laurent} retain the source and operator normalization.

With $\hat t=r_+t/L^2$, define the full impulse kernel and its action by
\begin{equation}
	K(\hat t)=\int_{\mathbb R+\iu0}\frac{\dd w}{2\pi}
	e^{-\iu w\hat t}G^{\mathrm R}_{\varkappa,{\rm FG}}(w),
	\qquad
	\langle O_-(\hat t)\rangle
	=\int\dd\hat t'K(\hat t-\hat t')J_\varkappa(\hat t').
	\label{eq:response_fourier}
\end{equation}
The corresponding kernel in physical time is
$(r_+/L^2)K(r_+t/L^2)$.  Equation~\eqref{eq:response_fourier} also makes the
driving dependence explicit: a source samples the full frequency response
through its spectrum, not only the nearest singularity.
The contribution of the double pole is instead
\begin{equation}
	K_{\rm pole}(\hat t)
	=-\iu\bigl(R_1^{\rm FG}-\iu R_2^{\rm FG}\hat t\bigr)
	e^{-\iu\wep\hat t}\Theta(\hat t).
	\label{eq:jordan_time_signal}
\end{equation}
Both Laurent terms belong to this contribution.  The factor linear in time is
the length-two Jordan signature; for $\operatorname{Im}\wep<0$ it remains a
damped relaxation \cite{Heiss:2012qla,Yang:2025dbn,Wang:2026robin}.  The
pole-splitting and coalescence checks are collected in the Supplemental
Material.
This fixes the four response objects used below.  $G^{\mathrm R}$ is the
frequency-domain boundary correlator, $K$ is its unfiltered inverse transform,
$K_{\rm pole}$ is the contribution of the displayed double pole, and
$K_\chi$ in eq.~\eqref{eq:filtered_response_scaling} will denote a frequency-windowed transform of $G^{\mathrm R}$.
Only the first and third are plotted in the main response figure.
Other poles and non-pole terms are present in $K$ and can affect the response
to a pulse even when the displayed pole controls the longest relaxation time.

\subsection{Peak height, bandwidth, amplitude, and duration}
\label{subsec:near_extremal}

We now ask how the frequency and impulse responses change as the co-rotating
EP becomes long lived.  Along the critical surface, the real-frequency peak
rises while its bandwidth shrinks; the double-pole impulse instead becomes
weaker while lasting longer.  The opposing peak trends are linked by the
shrinking frequency measure, and the duration follows from the same
near-extremal timescale.

\begin{figure}[t]
	\centering
	\btzpanel{0.48\textwidth}{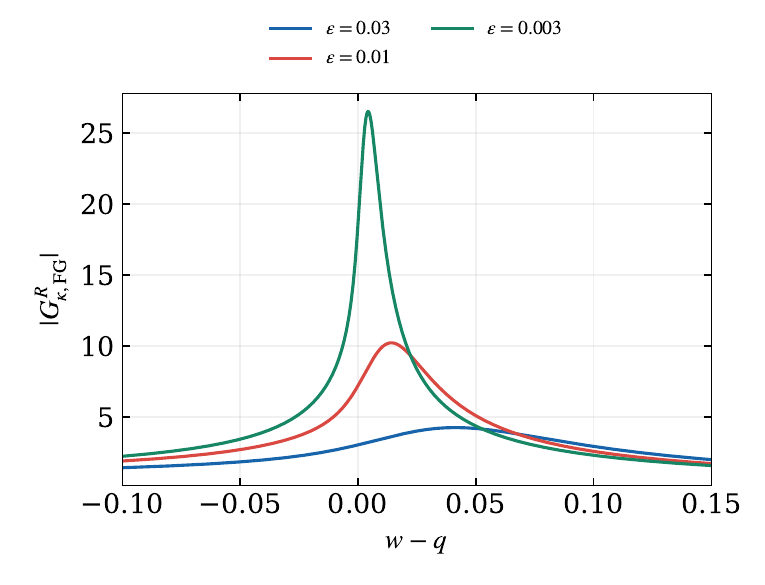}{Peak height grows as its frequency window narrows.}\hfill
	\btzpanel{0.48\textwidth}{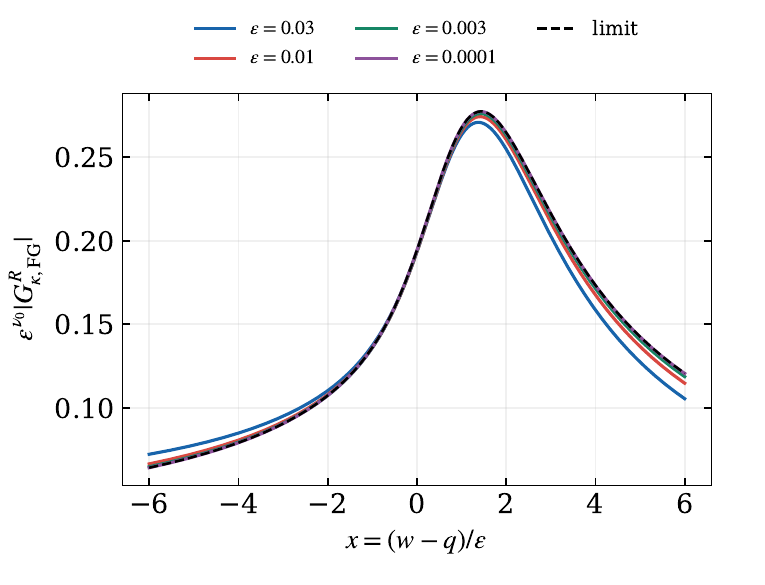}{Scaled correlators collapse onto $|\mathcal F|$.}
	\par\medskip
	\btzpanel{0.48\textwidth}{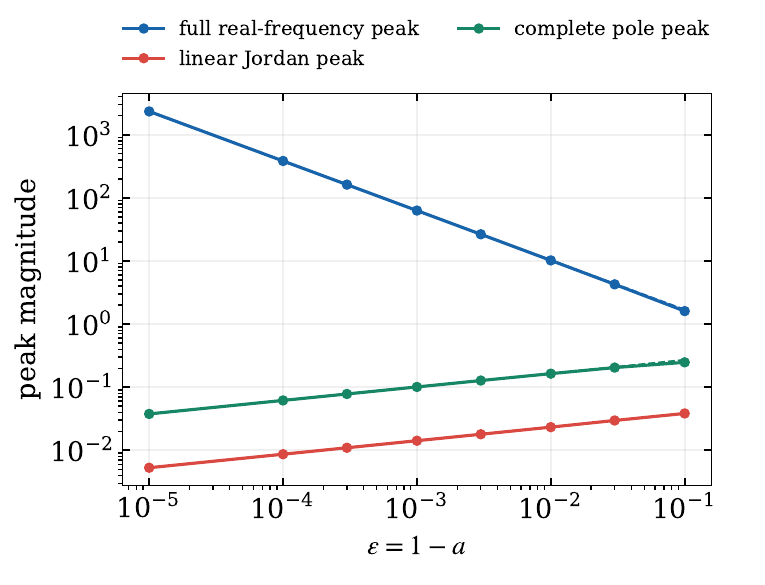}{Frequency and pole-kernel peaks have opposite powers.}\hfill
	\btzpanel{0.48\textwidth}{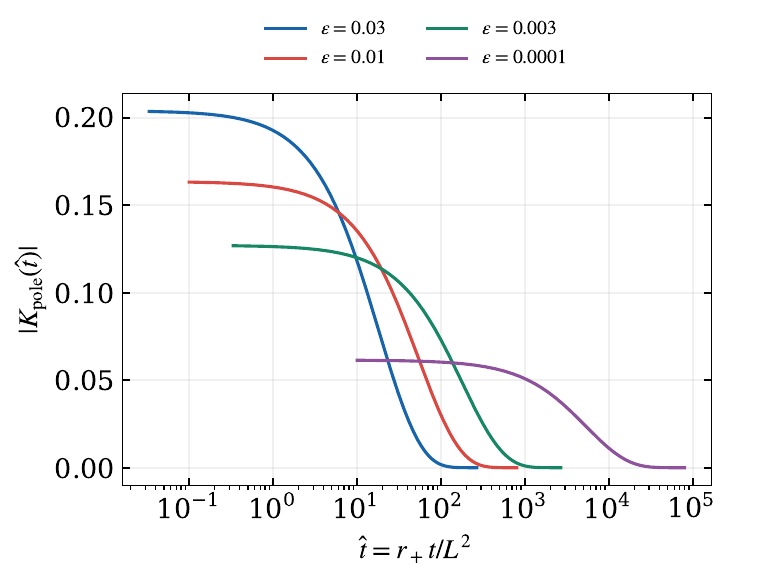}{The pole contribution weakens and lasts longer.}
	\caption{Near-extremal response on the co-rotating EP surface at $q=0.5$,
	with $m=L=\mu_R=1$ and $r_+=2$.
	Panels (a,b) evaluate the full FG correlator in fixed and scaled frequency
	windows.  Panel (c) compares its maximum with the peaks of the isolated
	Jordan term and $K_{\rm pole}$; the dashed slopes are $-\nu_0$ and
	$1-\nu_0$.  Panel (d) retains the damping and both Laurent terms in
	$K_{\rm pole}$.}
	\label{fig:near_extremal}
\end{figure}

We hold $q$ and $r_+$ fixed and vary both the scalar mass and boundary
coupling on the EP surface.  This differs from the fixed-theory crossing in
figure~\ref{fig:fixed_theory_slice}.  Set $\epsilon=1-a$ and resolve the
co-rotating window with $w=q+\epsilon x$.  For bounded real $x$ on compact
intervals where the limiting denominator is nonzero, let $(g_0,\nu_0)$ denote
the limiting critical controls and $A_0,B_0$ the $\epsilon\to0$ connection
coefficients.  The full FG correlator is
\begin{align}
	G^{\mathrm R}_{\varkappa,{\rm FG}}(q+\epsilon x)
	&=\epsilon^{-\nu_0}
	\left[\mathcal F(x)+O(\!\epsilon|\log\epsilon|)\right],
	\label{eq:full_real_frequency_scaling}\\
	\mathcal F(x)
	&=-2\nu_0(2r_+^2)^{-\nu_0}
	\frac{\Gamma(\nu_0)}{\Gamma(-\nu_0)}
	\frac{A_0(x;\nu_0)}{B_0(x;\nu_0)+g_0A_0(x;\nu_0)}.
	\label{eq:full_scaling_function}
\end{align}
Explicitly, $A_0$ and $B_0$ follow from
eqs.~\eqref{eq:A_coefficient}--\eqref{eq:B_coefficient} after taking
$x_{\co}\to x/2$, $x_{\ctr}\to q/2$, and $\nu\to\nu_0$.
Because $\nu_{\rm EP}=\nu_0+O(\epsilon)$, the FG conversion has the
asymptotic form
\begin{equation}
	Z_{\rm FG}=\epsilon^{-\nu_0}(2r_+^2)^{-\nu_0}
	\frac{\Gamma(\nu_0)}{\Gamma(-\nu_0)}
	[1+O(\!\epsilon|\log\epsilon|)].
	\label{eq:near_extremal_FG_normalization}
\end{equation}
The connection coefficients approach finite functions in the scaled window,
whereas $Z_{\rm FG}$ supplies the growing prefactor.  The variation of the
exponent accounts for the logarithmic correction.  This derivation uses the
exact correlator in eq.~\eqref{eq:retarded_response}; the Laurent coefficients
are not used.
Thus the peak height grows as $\epsilon^{-\nu_0}$ while its bandwidth in $w$
shrinks as $\epsilon$.  Figure~\ref{fig:near_extremal}(a,b) shows both changes
and the collapse onto $|\mathcal F|$.  The result is obtained from the full
correlator in the scaling window, rather than by extending the local Laurent
series to real frequencies.
The compact-$x$ qualification is essential: this limit resolves frequencies
whose distance from $q$ shrinks with the co-rotating temperature.  It does not
give a uniform approximation at fixed $w-q\neq0$ or over the entire real axis.
The width is defined by the two half-magnitude crossings of $|G^{\mathrm R}|$,
so it is a width of the response amplitude rather than of a squared spectral
density.  At $\epsilon=10^{-5}$, the complex scaled correlator agrees with
$\mathcal F$ to $4.3\times10^{-5}$ relative to the maximum of
$|\mathcal F|$ on $-6\leq x\leq6$.  The full samples and fitted powers are
tabulated in the Supplemental Material.

The height--bandwidth relation fixes the scale of a band-limited impulse.  Let
$K_\chi$ be the inverse transform after multiplying the
correlator by a smooth compact window $\chi((w-q)/\epsilon)$ within the regular
scaling interval.  Since $\dd w=\epsilon\,\dd x$,
\begin{equation}
	K_\chi(\hat t)
	=\epsilon^{1-\nu_0}e^{-\iu q\hat t}
	\int\frac{\dd x}{2\pi}\,\chi(x)\mathcal F(x)e^{-\iu xT}
	+o(\epsilon^{1-\nu_0}),
	\qquad T=\epsilon\hat t .
	\label{eq:filtered_response_scaling}
\end{equation}
The frequency peak can therefore rise while the corresponding band-limited
impulse decreases and spreads over $\hat t\sim\epsilon^{-1}$.  Equation
\eqref{eq:filtered_response_scaling} concerns $K_\chi$ and does not determine
the unfiltered kernel $K$.
The additional factor of $\epsilon$ is the shrinking measure of the frequency
window.  It is this height--bandwidth product, without assuming that the EP
pole dominates the real axis, that connects the first two panels of
figure~\ref{fig:near_extremal} to the impulse scale.
The carrier $e^{-\iu q\hat t}$ does not affect the envelope, while the scaled
time $T=\epsilon\hat t$ shows directly why a fixed shape in $T$ lasts longer
in the original time coordinate.

For complex frequencies write $w=q+\epsilon\zeta$.  The parameters entering
this limit come from
$x_{\co}=\zeta/2$ and
$x_{\ctr}=(2q+\epsilon\zeta)/[2(2-\epsilon)]$.  The resulting limiting
characteristic $\mathcal H_0=B_0+gA_0$ has a double root
$(\zeta_0,g_0,\nu_0)=(0.786691-1.894048\iu,0.324396,0.785662)$ at $q=0.5$,
and
\begin{equation}
	\wep=q+\epsilon\zeta_0+O(\epsilon^2),
	\qquad \gep=g_0+O(\epsilon),
	\qquad \nuep=\nu_0+O(\epsilon).
	\label{eq:near_extremal_analytic_law}
\end{equation}
The explicit $A_0,B_0$ construction and continuation checks are given in the
Supplemental Material.  Although $g$ approaches a nonzero constant, the FG
coupling vanishes:
\begin{equation}
	\widehat\varkappa_{\mu_R,{\rm EP}}
	=K_0\epsilon^{\nu_0}
	[1+O(\!\epsilon|\log\epsilon|)],
	\quad
	K_0=-(2\mu_R^2r_+^2)^{\nu_0}
	\frac{\Gamma(-\nu_0)}{\Gamma(\nu_0)}g_0>0.
	\label{eq:near_extremal_physical_coupling}
\end{equation}
The long-lived critical mode is therefore reached while the physical boundary
condition approaches the Neumann endpoint.
This correlated tuning of mass and boundary data is part of the critical path;
it cannot be replaced by holding the Gamma-normalized coordinate $g$ fixed and
calling that a fixed FG theory.
The imaginary part of eq.~\eqref{eq:near_extremal_analytic_law} also gives
$-\operatorname{Im}w_{\rm EP}=O(\!\epsilon)$, so the relaxation time is
$O(\!\epsilon^{-1})$ before any residue information is used.

The residue scalings reproduce the amplitude--duration balance on the same
timescale.  Differentiating with respect to
$w=q+\epsilon\zeta$ gives
\begin{equation}
	R_1^{\rm FG}=\epsilon^{1-\nu_0}
	[r_{1,0}+O(\!\epsilon|\log\epsilon|)],
	\qquad
	R_2^{\rm FG}=\epsilon^{2-\nu_0}
	[r_{2,0}+O(\!\epsilon|\log\epsilon|)],
	\label{eq:near_extremal_FG_residues}
\end{equation}
with finite nonzero $r_{1,0}$ and $r_{2,0}$.  Substitution into
eq.~\eqref{eq:jordan_time_signal} yields, for fixed $T>0$,
\begin{equation}
	K_{\rm pole}(\hat t)
	=-\iu\epsilon^{1-\nu_0}e^{-\iu q\hat t}
	\bigl(r_{1,0}-\iu r_{2,0}T\bigr)e^{-\iu\zeta_0T}
	+o(\epsilon^{1-\nu_0}).
	\label{eq:complete_pole_scaling}
\end{equation}
The $R_1$ exponential and $R_2$ Jordan term contribute at the same order.
The extra power of $\epsilon$ in $R_2^{\rm FG}$ is offset by
$\hat t=T/\epsilon$ on the relaxation timescale.
Their combined peak falls as $\epsilon^{1-\nu_0}$, while its duration grows as
$\epsilon^{-1}$.  The isolated Jordan term has the same peak power, although
its maximum need not coincide with the maximum of $K_{\rm pole}$.
For the Jordan term alone, the polynomial factor peaks against the damping at
$\hat t\sim[-\operatorname{Im}\wep]^{-1}$.  Including $R_1$ shifts the
maximum and changes the envelope, but not the leading power because both terms
are of order $\epsilon^{1-\nu_0}$ at fixed $T$.  Panels (c,d) verify this
combined behavior while retaining the exponential decay.
Because $0<\nu_0<1$, the amplitude tends to zero even as the decay time
diverges.  The long lifetime therefore does not compensate for the vanishing
pole weights on this path.

\subsection{What depends on the EP, and what depends on the path}
\label{subsec:response_controls}

\begin{figure}[t]
	\centering
	\btzpanel{0.48\textwidth}{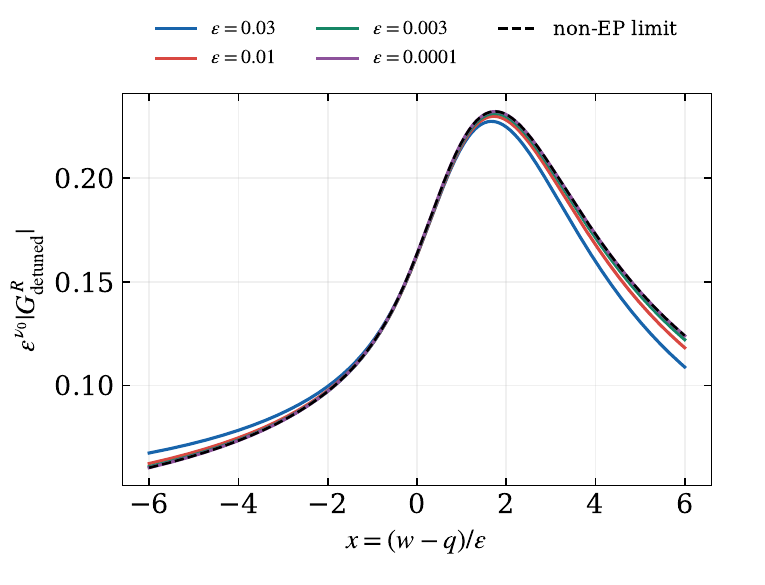}{Non-EP control: the frequency power persists.}\hfill
	\btzpanel{0.48\textwidth}{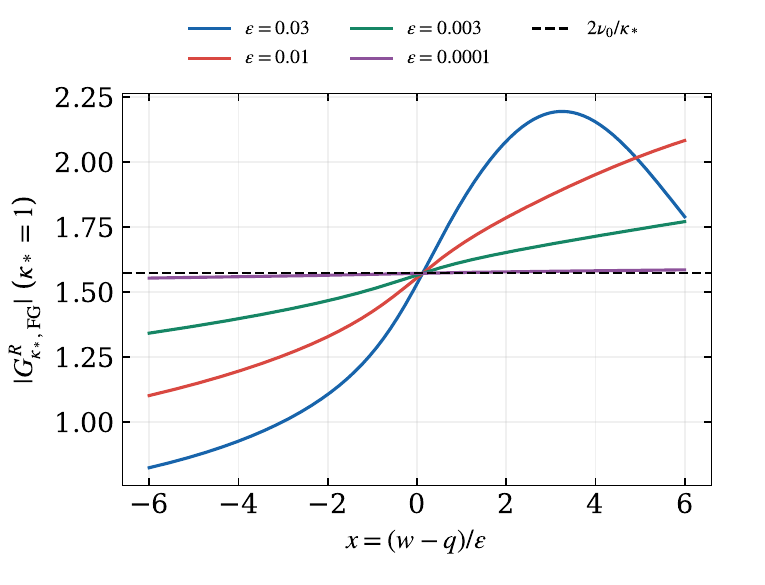}{Fixed nonzero FG coupling: a finite limit.}
	\caption{Controls of the near-extremal response.  Panel (a) uses
	$g=1.2g_{\rm EP}$ at the critical mass of each geometry; the dashed curve
	uses $1.2g_0$ in the limiting denominator.  Panel (b) fixes
	$\nu=\nu_0$ and $\varkappa_{\rm FG}=1$ while varying the state; its dashed
	line is $2\nu_0/\varkappa_{\rm FG}$.}
	\label{fig:near_extremal_controls}
\end{figure}

The two controls in figure~\ref{fig:near_extremal_controls} separate the
double-root physics from the path to extremality.  First, detuning to
$g=1.2g_{\rm EP}$ at the same critical mass splits the double root but retains
the $\epsilon^{-\nu_0}$ frequency power, with a different limiting line shape.
The double root is therefore unnecessary for this power; it remains essential
for the double pole and the Jordan factor in eq.~\eqref{eq:jordan_time_signal}.
The control tests whether the double root is necessary for the exponent; the
limiting line shapes, pole decompositions, and time-domain excitation remain
different.

Second, fixing $\nu=\nu_0$ and a nonzero
$\varkappa_*=\varkappa_{\rm FG}$ changes the physical path.  Because
$g=-\varkappa_*Z_{\rm FG}$,
\begin{equation}
	G^{\mathrm R}_{\varkappa_*,{\rm FG}}(q+\epsilon x)
	=\frac{2\nu}{\varkappa_*}
	\left[1-\frac{B/A}{\varkappa_*Z_{\rm FG}}\right]^{-1}
	\longrightarrow\frac{2\nu}{\varkappa_*}.
	\label{eq:fixed_theory_response_limit}
\end{equation}
The limit is finite on the same regular compact $x$ intervals and in the
prescription fixed above.  It shows that the vanishing FG coupling along the
critical surface participates in the growing frequency response.  Contact
polynomials with coefficients bounded in the limit can shift a finite result
but cannot change the divergent critical-path power; singular scheme changes
are outside this comparison.

The protocol determines which response measure is sampled.  For a
monochromatic FG source of fixed amplitude $J_0$, the steady-state response at
$w_d=q+\epsilon x$ has magnitude $|G^{\mathrm R}(w_d)J_0|$.  The drive must
resolve the $O(\!\epsilon)$ window and persist beyond the
$O(\!\epsilon^{-1})$ relaxation time.  A pulse instead samples the convolution in
eq.~\eqref{eq:response_fourier}.  The steady-state enhancement and the
decreasing $K_{\rm pole}$ amplitude therefore apply to distinct driving
protocols; no response for an arbitrary source is inferred here.
All statements are within linear response; a fixed nonzero source can
eventually leave that regime as the susceptibility grows.

\subsection{Exact finite-duration drive}
\label{subsec:finite_drive}

Fix a radial-amplitude unit $R_*$ and define
$j=\mu_R^{1+\nu}J_\varkappa/R_*$ and
$o=\mu_R^{1-\nu}\langle O_-\rangle/R_*$.  Thus, in frequency space,
$\widetilde o(w)/\widetilde j(w)
=\mu_R^{-2\nu}G^{\rm R}_{\varkappa,{\rm FG}}(w)$; here
$\mu_R=R_*=1$.  Comparisons with
different $\nu$ use this common engineering convention for different critical
theories.  To place the frequency-domain result in a specified driving
protocol, consider
\begin{equation}
 j(\hat t)=j_0s(\hat t/\tau_d)e^{-\iu w_d\hat t},\qquad
 w_d=q+\epsilon x_d,\qquad \eta=\epsilon\tau_d ,
 \label{eq:finite_drive_source}
\end{equation}
where the same peak-one $C^\infty$ envelope $s$ has 20\% smooth ramps and a
plateau on $0.2\leq\hat t/\tau_d\leq0.8$.  If
$\widehat s(y)=\int_0^1s(u)e^{\iu yu}\dd u$, then
$\widetilde j/j_0=\tau_d\widehat s[\tau_d(w-w_d)]$ and the exact response is
\begin{equation}
 \frac{o(\hat t)}{j_0}
 =\eta e^{-\iu q\hat t}\int\frac{\dd x}{2\pi}\,
 e^{-\iu xT}G^{\rm R}(q+\epsilon x)
 \widehat s[\eta(x-x_d)],\qquad T=\epsilon\hat t .
 \label{eq:finite_drive_exact}
\end{equation}
This is a full-frequency convolution, rather than an application of
eq.~\eqref{eq:complete_pole_scaling} or the windowed kernel in
eq.~\eqref{eq:filtered_response_scaling}.

\begin{figure}[t]
 \centering
 \includegraphics[width=0.98\textwidth]{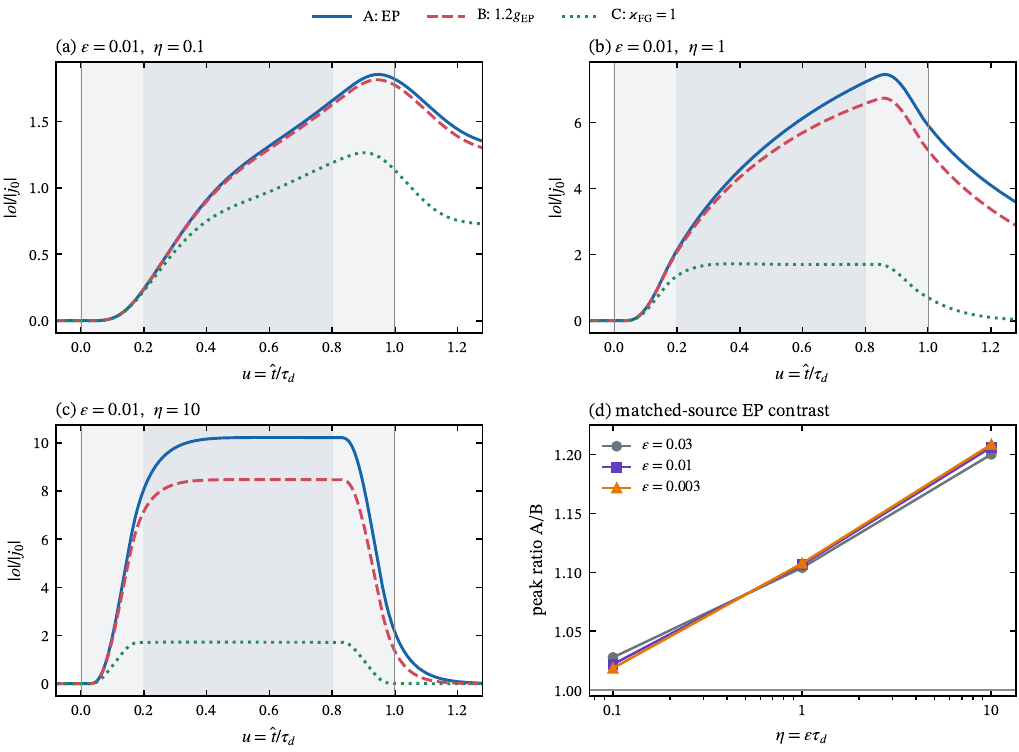}
 \caption{Matched finite-duration response from the exact correlator.
 Panels (a--c) show $|o|/|j_0|$ at $\epsilon=0.01$ against
 $u=\hat t/\tau_d$; light shading marks the source support and darker shading
 its plateau.  Each panel includes only the early post-drive segment in its
 own relative-duration coordinate, so panel (a) is not a plot of the complete
 relaxation.  Panel (d) summarizes the A/B peak ratio for the three computed
 $\eta$ values; connecting lines are guides to the eye.}
 \label{fig:finite_drive_response}
\end{figure}

We fix $q=0.5$, $x_d=1.414858$, $j_0=1$, and use
$\epsilon=0.03,0.01,0.003$ with $\eta=0.1,1,10$.  Within every
$(\epsilon,\eta)$ cell the EP path A, the matched non-EP path
$g=1.2g_{\rm EP}$ (B), and the fixed-coupling path
$\nu=\nu_0$, $\varkappa_{\rm FG}=1$ (C) receive identical sources.
Figure~\ref{fig:finite_drive_response} shows that, at $\epsilon=0.01$, the
A/B peak ratio grows from $1.022$ for $\eta=0.1$ to $1.206$ for
$\eta=10$.  Moreover, $|G_B(w_d)|$ lies within $1.2\%$ of
B's own frequency peak, so the contrast is not explained by appreciable
detuning of B.  For $\eta=10$, the drive plateau agrees with
$G(w_d)j_0$ to within $0.18\%$, and the residual response decays over the
checked interval after the source is switched off.

Because the source spectrum contains $\tau_d=\eta/\epsilon$, a fixed-$\eta$
drive does not inherit the $\epsilon^{1-\nu_0}$ amplitude of the windowed
impulse or of $K_{\rm pole}$.  The exact finite-$\epsilon$ trends are
compatible with the fixed-$\eta$ critical-window scale analysis, but are not a
full-frequency uniform $\epsilon\to0$ proof or a new asymptotic power.  A
separate fixed-duration check, with the carrier changed by the same fixed-$x_d$
rule, is reported in the Supplemental Material.  Recorded post-drive samples
extend to $T-\eta=8$, but only support a finite-interval shape comparison.
Source norms and numerical convergence tests are also given there.

\section{A regulated resolvent diagnostic}
\label{sec:pseudospectrum}

The boundary correlator fixes one source--response channel.  As a separate
diagnostic of the defect, we construct a finite regulated bulk evolution
problem
\begin{equation}
	\iu\partial_\tau\mathbf u=\mathcal L\mathbf u,
	\qquad \mathbf u=(\psi,\partial_\tau\psi)^{\mathsf T},
	\qquad \|\mathbf u\|_E^2=\mathbf u^\dagger G_E\mathbf u,
	\quad G_E=F^\dagger F.
	\label{eq:first_order_generator}
\end{equation}
The horizon-corotating hyperboloidal slices are ingoing at the future horizon
and meet a finite outer cutoff $s_c=r_+/r_c$ without a time tilt.  A real,
frequency-independent Robin condition fixes the operator domain, and the
regulated Klein--Gordon energy includes its boundary term.  These choices are
held fixed throughout each frequency scan.  In energy-normalized coordinates,
\begin{equation}
	\widehat{\mathcal L}_E=\frac{L^2}{r_+}
	\left(F\mathcal L F^{-1}+m\Omega_H I\right),
	\qquad
	s_{\min,E}(w)=s_{\min}(wI-\widehat{\mathcal L}_E).
	\label{eq:energy_similarity}
\end{equation}
The inverse of $s_{\min,E}$ is the resolvent norm for this finite operator and
energy metric.  A small value therefore measures sensitivity to unrestricted
forcing in the regulated bulk state norm; it is not the gain of the FG
boundary channel defined in section~\ref{sec:holography}.

At a size-two Jordan defect, $s_{\min,E}\propto r^2$ as the distance
$r=|w-w_c|$ tends to zero; near an isolated simple pole it is linear.  We tune
the regulated generator itself to a strict EP using an independent bordered
characteristic and Jordan-vector test.  At $(s_c,N)=(0.002,44)$, its angular
median follows the quadratic law, whereas a far-detuned simple pole is linear.
Construction, certification, fitted powers, precision residuals, and
cutoff/order checks are provided in the Supplemental Material.

A nearby pair of simple poles explains why an apparently quadratic regime is
not sufficient to certify defectiveness.  For $g=g_c+10^{-6}$, let
$\Delta=|w_+-w_-|\simeq1.1\times10^{-3}$ and center the circles on one pole.
The measured regimes are
\begin{equation}
	\operatorname{median}_\theta s_{\min,E}
	\propto
	\begin{cases}
		r, & 0.01\leq r/\Delta\leq0.15,\\
		r^2, & 5\leq r/\Delta\leq20.
	\end{cases}
	\label{eq:energy_crossover_orders}
\end{equation}
Below the splitting scale the two poles are resolved and each has a linear
neighborhood.  Above it, the unresolved pair mimics the quadratic behavior of
an EP over an intermediate range.  A strict defect differs in retaining the
quadratic law as $r\to0$.

\begin{figure}[t]
	\centering
	\includegraphics[width=0.62\textwidth]{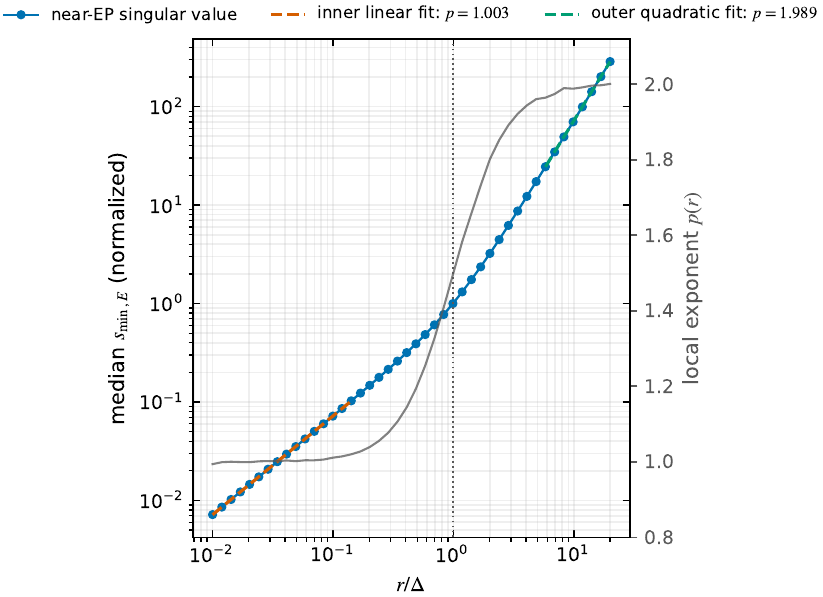}
	\caption{Regulated KG-energy diagnostic for a close simple-pole doublet.
	Circles are centered on one pole.  The lower and upper scale bars mark the
	linear $r\ll\Delta$ and approximately quadratic $r\gg\Delta$ regimes.}
	\label{fig:energy_pseudospectrum}
\end{figure}

This calculation establishes a distinction between a strict finite-matrix EP
and an unresolved doublet for the stated regulated operators.  Moving the
cutoff brings the critical spectral location toward the exact characteristic,
but also changes the foliation and energy Gram matrix.  We therefore do not
infer a continuous-limit pseudospectrum or a universal amplification in a
particular boundary channel.

\section{Conclusions}
\label{sec:conclusions}

Rotation separates the critical mass and Robin data of the two chiral BTZ EP
surfaces.  Their large raw damping contrast is mainly set by the chiral thermal
scales, although a resolved residual remains after thermal normalization.  A
critically selected path also shows that varying the black-hole spin can cross
the defect while the scalar data and FG coupling are held fixed.

The main response result follows the co-rotating critical surface toward
extremality, where the FG coupling tends to zero.  In the scaled real-frequency
window, the correlator peak rises as $\epsilon^{-\nu_0}$ while its bandwidth
shrinks as $\epsilon$.  The frequency measure then makes the band-limited
impulse scale as $\epsilon^{1-\nu_0}$.  The same amplitude power follows for
$K_{\rm pole}$ because its ordinary and Jordan Laurent terms contribute
together on the $\epsilon^{-1}$ relaxation timescale.  The resulting pole
contribution is weaker and longer lived.

The controls locate the origin of these features.  Detuning the double root
preserves the frequency power but removes the EP pole and Jordan time
structure.  Holding a nonzero FG coupling instead produces a finite limit on
the regular scaled window.  A persistent frequency-matched drive can probe the
growing susceptibility; a pulse depends on its spectrum and the full
convolution.  Long lifetime therefore does not determine response magnitude
without a physical path, normalization, and drive.
For the compact smooth source tested here, exact full-frequency convolution
gives an EP/non-EP peak ratio from about $1.02$ for a short drive to about
$1.20$ for a long drive.  The finite-$\epsilon$ trends are compatible with
the critical-window scale analysis, but do not define another asymptotic
exponent or an EP-specific amplification law.

Finally, the exact double zero and one-dimensional ingoing space establish the
geometric defect.  The regulated energy resolvent gives a complementary
finite-matrix test: a strict EP remains quadratic as $r\to0$, while a close
simple-pole pair becomes linear below its splitting scale.  This does not
establish a continuous-limit pseudospectrum or universal observable
amplification.

\acknowledgments
This work is supported by the National Natural Science Foundation of China
(Grants Nos. 12547143, 12305063, 12275079, 12405055 
, 2035005, 12475049 and 12447156), China Postdoctoral
Science Foundation (Grant No. 2025M773339), the National Key Research and
Development Program of China (Grant No. 2020YFC2201400 and 2021YFC2203001), the Key Project of the Department of Education of Hunan Province (No.
25A0084), the Hunan Provincial Furong Plan Youth Talent Support Program (Grant No. 2026ZRB-QN100) and the innovative
research group of Hunan Province (Grant No. 2024JJ1006).

\bibliographystyle{JHEP}
\bibliography{references}

\end{document}